\documentclass[twocolumn]{aastex62}
\graphicspath{{./}{figures/}}

\shorttitle{No evidence of the grain growth in the protostellar disk around L1489}
\shortauthors{Ohashi et al.}

\begin{document}

\title{No evidence of the significant grain growth but tentative discovery of disk substructure in a disk around the Class I Protostar L1489 IRS}

\author[0000-0002-9661-7958]{Satoshi Ohashi}
\affil{RIKEN Cluster for Pioneering Research, 2-1, Hirosawa, Wako-shi, Saitama 351-0198, Japan}
\email{satoshi.ohashi@riken.jp}

\author[0000-0001-8808-2132]{Hiroshi Kobayashi}
\affil{Department of Physics, Graduate School of Science, Nagoya University, Furo-cho, Chikusa-ku, Nagoya 464-8602, Japan}

\author[0000-0003-4361-5577]{Jinshi Sai}
\affiliation{Department of Astronomy, Graduate School of Science, The University of Tokyo, 7-3-1 Hongo, Bunkyo-ku, Tokyo 113-0033, Japan}
\affiliation{Academia Sinica Institute of Astronomy and Astrophysics, 11F of Astro-Math Bldg, 1, Sec. 4, Roosevelt Road, Taipei 10617, Taiwan}

\author[0000-0002-3297-4497]{Nami Sakai}
\affil{RIKEN Cluster for Pioneering Research, 2-1, Hirosawa, Wako-shi, Saitama 351-0198, Japan}



\begin{abstract}

For revealing the first step of the plant formation, it is important to understand how and when dust grains become larger in a disk around a protostar.
To investigate the grain growth, we analyze dust continuum emission toward a disk around the Class I protostar, L1489 IRS at 0.9 and 1.3 mm wavelengths obtained by the Atacama Large Millimeter/submillimeter Array.
The dust continuum emission extends to a disk radius ($r$) of $r\sim300$ au, and  the spectral index ($\alpha$) is derived to be $\alpha\sim3.6$ at a radius of $r\sim100-300$ au, as similar to the interstellar dust. Therefore, the grain growth does not occur significantly in the outer disk ($r\sim100-300$ au).
Furthermore, we tentatively identify a ring-like substructure at $r\sim90$ au even though the spatial resolution and sensitivity are not enough to determine this structure.
If this is the real ring structure, the ring position and small dust in the disk outer part are consistent with the idea of the growth front.
These results suggest that the L1489 protostellar disk may be the beginning of the planet formation.

\end{abstract}



\section{Introduction} \label{sec:intro}

The first step of the planet formation is suggested to begin by grain growth via coagulation of the initial dust of the interstellar medium (ISM) in a protoplanetary disk.
Recent  high-resolution observations with Atacama Large Millimeter/submillimeter Array (ALMA) have found multiple gaps and rings in protoplanetary disks at submillimeter wavelengths \citep{alma15,and18}.
The presence of large grains (millimeter sized) in protoplanetary disks have been studied by spatially integrated measurements of spectral indices from the submillimeter to centimeter wavelength range with multi-wavelength observations \citep[e.g.,][]{tsu16,van18,car19,mac19,lon20,hua20,pan21}.
Furthermore, existences of planets have been pointed out in several protoplanetary disks \citep{pin18,tea18,alv20}.
These results suggest that the planet formation may start earlier than classical theories.

Interestingly, dust ring structures are observed not only in Class II protoplanetary disks, but also even in earlier stages of disk formation around Class 0 and I objects \citep{alma15,seg20,she20}.
Even though it can be speculated that the ring structures imply the grain growth even in the early stage of the protostellar disks, the grain size distribution has been unclear because the dust emission is highly optically thick or only single band observations have been carried out in the protostellar disks.
The measurement of the dust size have been studied toward protostellar disks regardless of substructures via dust continuum emission. Several observations have suggested that the dust grains have already significantly grown at the Class 0/I stage \citep[e.g.,][]{kwo09,zha15,cie16,har18} even though these observations may be affected by the high optical depth.
In contrast, recent observations toward a protostellar disk of IRAS16293-2422 B have suggested that the dust grains may not become larger ($a_{\rm max}\leq 10$ $\mu$m) \citep{zam21}.
Therefore, it is important to understand how and when the dust grains are grown by revealing spatial distribution of both the initial ISMs dust and larger dust grains in protostellar disks with optically thin dust emission.

A L1489 IRS protostar is located in the Taurus molecular cloud at a distance of 140 pc measured by Gaia \citep{zuc19} and is classified as a Class I protostar \citep{fur08}. A large-scale outflow and surrounding envelope are associated with this protostar \citep{yen14}.  The conditions of the deeply embedded protostar associated with the outflow represent an early stage of the star-forming process. A Keplerian rotation associated with the L1489 protostar was identified by C$^{18}$O molecular line observations with a disk radius of 600 au \citep{bri07,yen13,yen14,sai20}. An infalling motion from the envelope toward the Keplerian disk is also identified, indicating that the disk is still growing \citep{yen14}. 
In this paper, we report multi-frequency observations of the protostellar disk around L1489 IRS with ALMA to probe the detailed disk structure and investigate dust spectral index.

\section{Observations} \label{sec:obs}

We have retrieved the archival ALMA observations toward L1489 IRS for the continuum emission at Band 7 (2015.1.01549.S) and Band 6 (2013.1.01086.S).  These data are selected because of better sensitivity and spatial resolutions than others.
The wavelengths of the continuum emission are 0.9 mm (330 GHz) and 1.3 mm (234 GHz), respectively.

For the 0.9 mm dust continuum emission, the observations were carried out with a configuration of the baseline length from 15.1 m to 1124.3 m on 2016 July 26. The maximum recoverable scale ($\theta_{\rm MRS}$) is estimated as $\sim 0.6 \lambda/L_{\rm min}$, where $\lambda$ is the observing wavelength and $L_{\rm min}$, is the minimum baseline. Since $\theta_{\rm MRS}$ is estimated to be $\sim7\farcs1$ (corresponding to $\sim1000$ au), any structures extended more than that size will be resolved out. The phase center was  ($\alpha(2000), \delta(2000)$) = (4h4min42.85s, 26$^\circ$18$'$56$\farcs30$). The on-source times are $\sim15$ minutes. The calibration of the data was done with CASA 4.5.3. The flux scale is calibrated with J0510+1800 have an uncertainty of $\sim10\%$ \citep{lun13}. In addition, to improve the sensitivity and image fidelity, self-calibration only for phase was performed.
The interval time to solve the complex gain was set to be infinity.

For the 1.3 mm dust continuum emission, the observations were carried out with two configurations: one with the baseline length from 20.6 m to 558.2 m on 2015 May 24 and the other extended one with the baseline length from 40.6 m to 1507.9 m on 2015 September 20. The maximum recoverable scale ($\theta_{\rm MRS}$) is $7\farcs8$ (corresponding to $\sim1100$ au). The phase center during the two tracks was  ($\alpha(2000), \delta(2000)$) = (4h4min42.8s, 26$^\circ$18$'$56$\farcs30$). The on-source times are $\sim23$ minutes and $\sim25$ minutes for the former and latter observations, respectively. The calibration of the data with longer and shorter baselines was done with CASA 4.5.0 and CASA 4.5.2, respectively. The flux scales are calibrated with J0510+180 and J0423-013 and have an uncertainty of $\sim10\%$ \citep{lun13}. In addition, to improve the sensitivity and image fidelity, self-calibration only for phase was performed.
The interval time to solve the complex gain was  set to be infinity.

The imaging is performed in CASA task tclean with CASA version 5. 
We construct two images for the both Band 7 and 6 data; a coarser spatial resolution with a better sensitivity and a better spatial resolution with a worse sensitivity to investigate the dust thermal emission for a wide range of the spatial scale of the disk structure.
For the images with the larger beam size and better sensitivity, the Natural weighting with the uvtpaper of 0.25 arcsec  (corresponding to 50 k$\lambda$ in the uv range) was used to improve the sensitivity. The synthesized beam sizes result in $0\farcs377\times0\farcs312$ with a position angle (PA) of $32^\circ$ and $0\farcs408\times0\farcs344$ with a PA of  $38^\circ$ in the 1.3 mm and 0.9 mm data, respectively. 
For the images with the smaller beam size, the Briggs weighting with a robust value of 0.5 and with a robust value of  $-2$ was used for the 1.3 mm and 0.9 mm data, respectively. 
The synthesized beam sizes result in $0\farcs240\times0\farcs158$ with a position angle (PA) of $32^\circ$ and $0\farcs263\times0\farcs176$ with a PA of $-26^\circ$ in the 1.3 mm and 0.9 mm data, respectively.

The spectral index between the 1.3 mm and 0.9 mm emission is derived in the similar uv rage by using CASA task tclean with the multi-frequency synthesis (MFS) method \citep[$nterm=2$,][]{rau11}. 
We confirmed that the continuum peak positions are consistent between the 1.3 and 0.9 mm  images before deriving the spectral index.
By applying different uv weightings in the tclean imaging, we obtain two images of the spectral index with the different spatial resolutions and sensitivities. The large-scale image with a better sensitivity is obtained by applying the Natural weighting and uvtaper of 0.25 arcsec (corresponding to 50 k$\lambda$ in the uv range), which allows us to derive the spectral index in a large scale of $r\sim300$ au. In contrast, the small-scale image with a better spatial resolution is obtained by applying the Briggs weighting with a robust value of 0.5, which allows us to resolve the detail structure of the disk. The synthesized beam sizes result in  $0\farcs368\times0\farcs338$ with a position angle (PA) of $19^\circ$ and $0\farcs227\times0\farcs179$ with a PA of $19^\circ$, respectively.

\section{Results} \label{sec:res}

\begin{figure*}[htbp]
\begin{center}
\includegraphics[width=16.cm,bb=0 0 3555 2262]{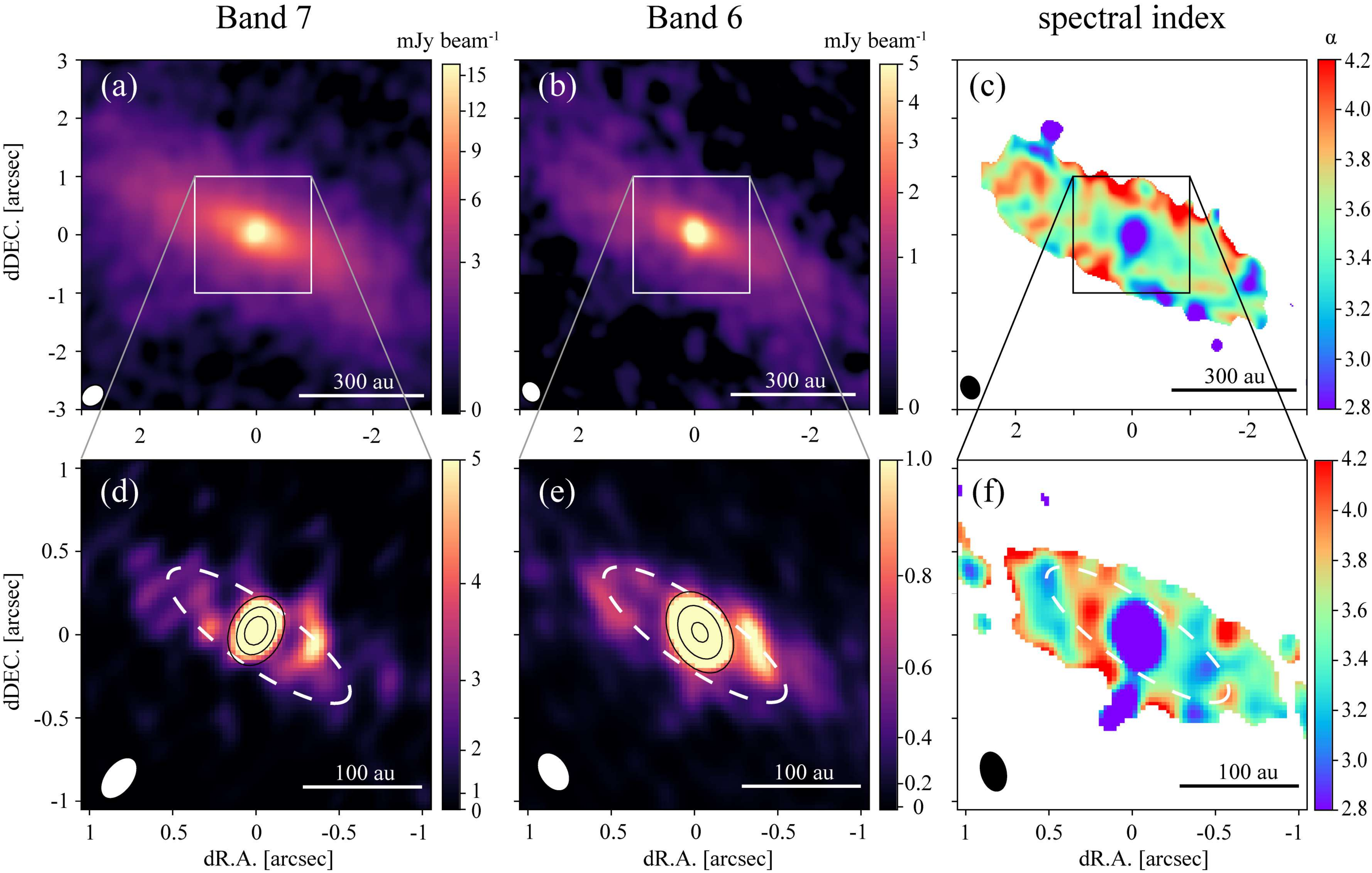}
\end{center}
\caption{Images of the disk around the L1489 IRS protostar at wavelengths of 0.9 mm (a) and 1.3 mm (b) with the larger beam size and better sensitivity.  The RMS noise level is $0.18$ mJy beam$^{-1}$ and $0.077$ mJy beam$^{-1}$, respectively. The spectral index map (c) is derived by these emission. Images of the 0.9 mm (d) and 1.3 mm (e) dust continuum emission with the smaller beam size are also shown. Emission from the central position is also shown as the black contours. The contours are 3, 6, 9 mJy beam$^{-1}$ in the panel (d) and 0.5, 2, and 5 mJy beam$^{-1}$ in the panel (e), respectively. The images indicate the ring structure (marked by the whited dashed ellipse) in the protostellar disk. The RMS noise level is $0.54$ mJy beam$^{-1}$ and $0.07$ mJy beam$^{-1}$, respectively. The spectral index map is also shown in  (f). 
In each panel the resolution is shown by a white or black ellipse in the lower left, a scale bar is in the lower right.
}
\label{fig1}
\end{figure*}

Figure \ref{fig1} (a) and (b) show intensity maps with the larger beam sizes at ALMA Band 7 and 6, respectively.
The rms noise level is $0.18$ mJy beam$^{-1}$ and $0.077$ mJy beam$^{-1}$, respectively.
Both images show disk structure extending over $r>300$ au.
The total flux densities are $290\pm12$ and $57\pm1.7$ mJy for Bands 7 and 6, respectively measured by integrating the flux in the disk structure on the image planes.

Figure \ref{fig1} (c) shows a spectral index $\alpha$ map (see Equation (\ref{eq1}) for its definition). The distribution seems to be approximately constant with $\alpha\sim3.6-3.8$ excepting for the disk center, and the $\alpha$ value of $\alpha\sim3.6-3.8$ is consistent with that observed in the ISMs \citep[e.g.,][]{dra06}.
Therefore, the dust size of the disk may be similar to the ISM-like dust.
We discuss the dependence of $\alpha$ on the dust size and grain growth in the disk in Section 4.1.

\begin{figure}[htbp]
\begin{center}
\includegraphics[width=8.cm,bb=0 0 2061 2556]{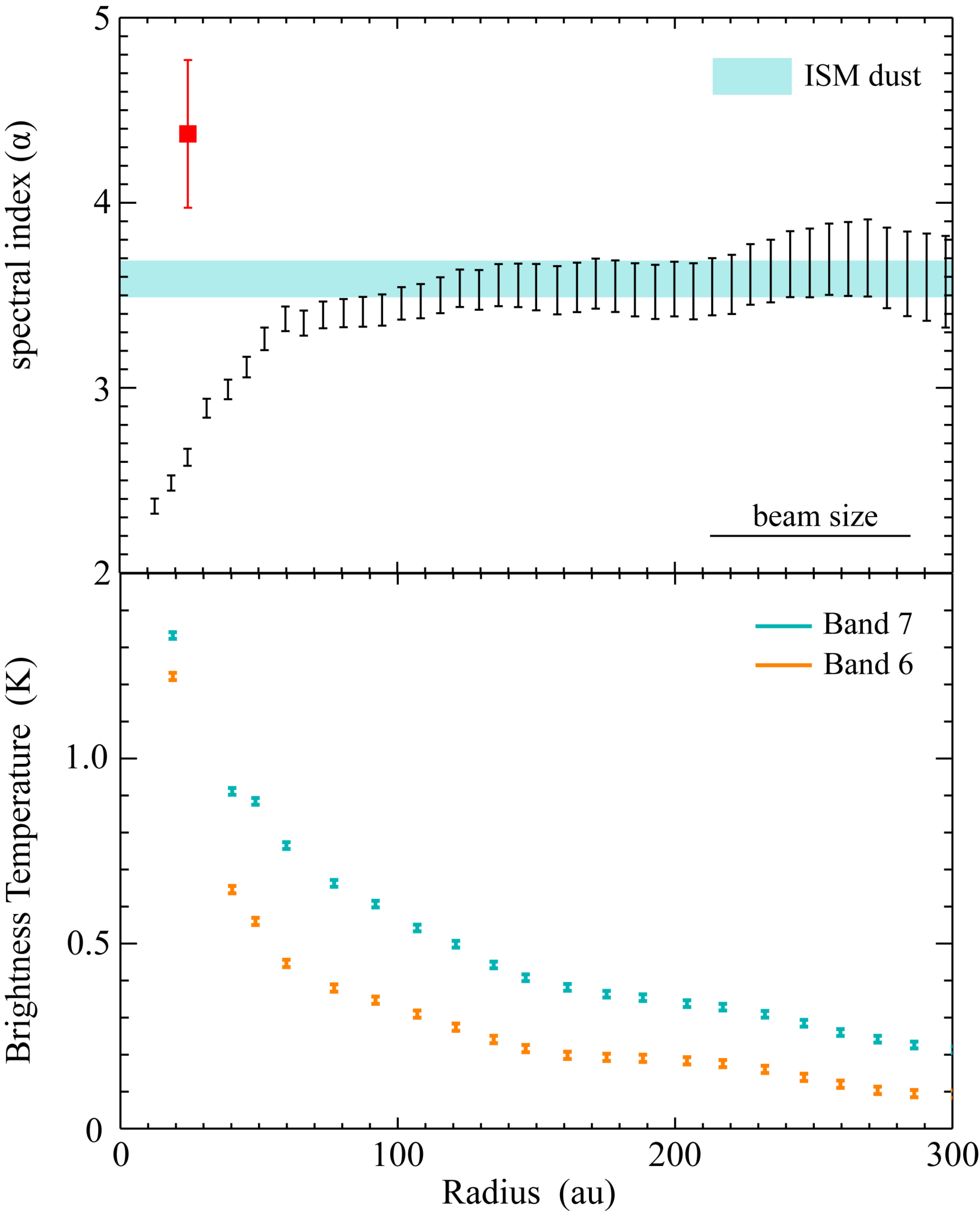}
\end{center}
\caption{The upper panel shows the radial plot of the spectral index $\alpha$ measured at 0.9 and 1.3 mm wavelengths.
The dashed horizontal line represents $\alpha=3.68$ for indicating the grain size smaller than $a_{\rm max}\lesssim100$ $\mu$m. 
The lower panel shows the radial plot of the intensities of the 0.9 and 1.3 mm dust thermal emission.
The radial profiles are made by averaging the northeast and southwest directions with position angles of $50-70^\circ$ and $230-250^\circ$, which are the major axis of the disk.
The 1$\sigma$ error bars of the spectral index in Figure \ref{fig2} are derived from the image noise levels. In addition, the impact of a flux calibration
error of 10\% is shown by the red error bar.
}
\label{fig2}
\end{figure}

We plot the radial profiles of the brightness temperatures of Band 7 and 6, and that of $\alpha$ in Figure \ref{fig2}.
The radial profiles are made by averaging the northeast and southwest directions with position angles of $50-70^\circ$ and $230-250^\circ$, which are the major axis of the disk.
We confirm that the spectral index shows a constant value of $\alpha\sim3.6\pm0.4$ at $r\gtrsim100$ au.
In contrast, the rapid decrease of $\alpha$ is found inside 50 au, which can be explained either by increase of optical depth or an existence of larger dust grains.
We cannot determine the origin of the $\alpha$  decrease in the center because it is difficult to measure the optical depths of these emission with the current large beam sizes.
The 1$\sigma$ error bars of the spectral index in Figure \ref{fig2} are derived from the image noise levels. In addition, the impact of a flux calibration
error of 10\% is shown by the red error bar.
The 10\%  accuracy of the absolute flux calibration causes an error of $\Delta\alpha\sim0.4$.
Note that this error due to the flux uncertainty does not affect spatial variations in the $\alpha$ value, but only the overall vertical shift.

The brightness temperatures of Band 7 and 6 show as low as $\lesssim1.0$ K at $r\gtrsim100$ au, which is much lower than the disk temperature of $\sim30$ K suggested by \citet{bri07}.
Such low brightness temperature indicates that the emission is optically thin.
This can also be supported by the spectral index much larger than 2.0.

\begin{figure*}[htbp]
\begin{center}
\includegraphics[width=16.cm,bb=0 0 2080 1531]{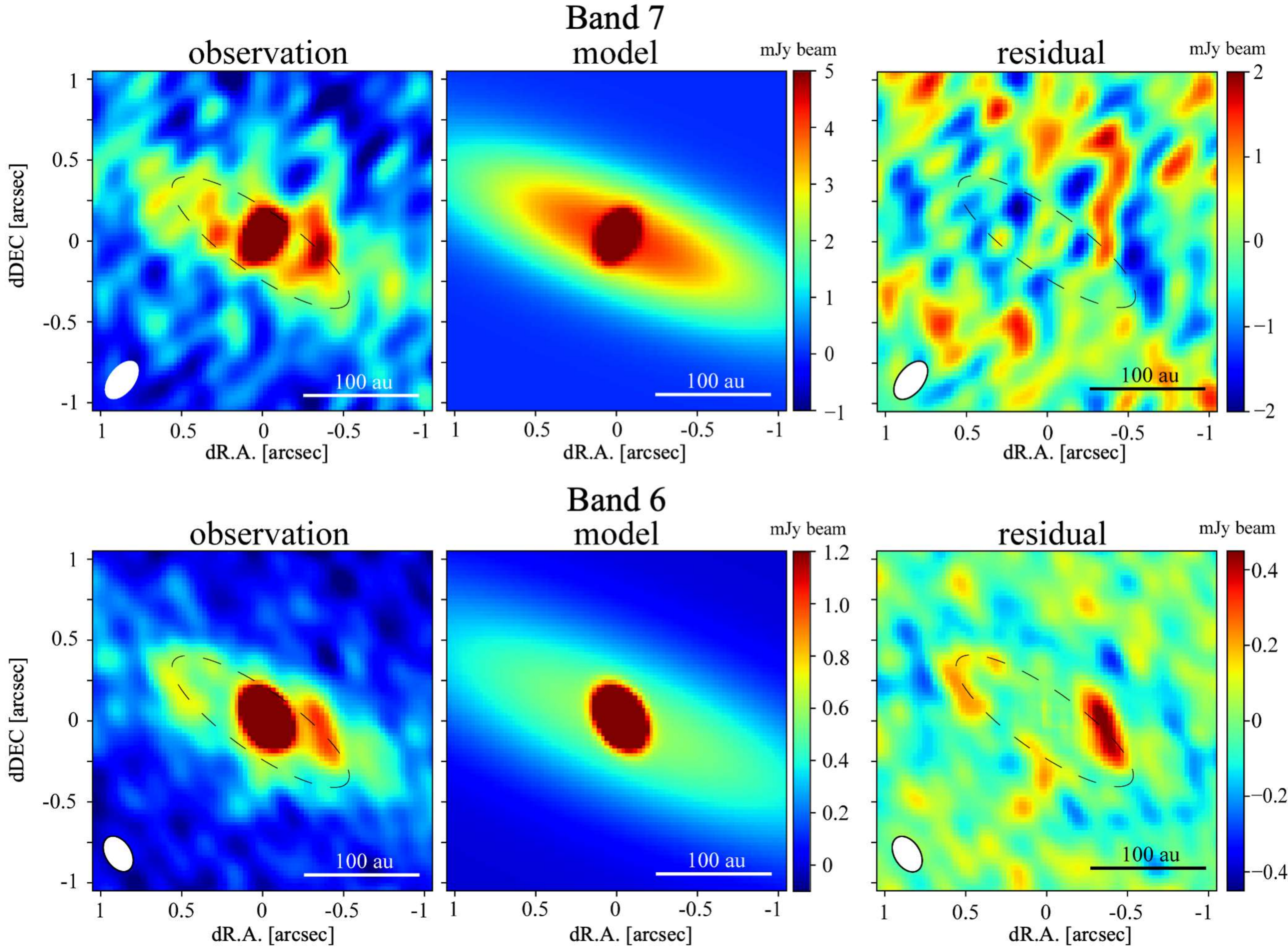}
\end{center}
\caption{ Images of the 0.9 mm and 1.3 mm dust continuum emission as the same with Figures \ref{fig1} (d) and (e) but the central components are subtracted. The RMS noise level is $0.54$ mJy beam$^{-1}$ and $0.07$ mJy beam$^{-1}$, respectively. The contours and the beam sizes are the same with Figure \ref{fig1} (d) and (e), respectively.
}
\label{residual}
\end{figure*}

Figure \ref{fig1} (d) and (e) show the zoomed-in views around the disk center with the smaller beam sizes of Band 7 and 6, respectively.
 The rms noise level is $0.54$ mJy beam$^{-1}$ and $0.07$ mJy beam$^{-1}$, respectively.
 Even though the beam sizes and sensitivities of these emission are not enough to reveal the morphology of the disk, the both images seem to show ring-like substructure at $r\sim 90$ au.
The white dashed circles represent the ring structure as eye guide.

To investigate the ring-like substructures in detail, we fit the images of Figures \ref{fig1} (d) and (e)  with two-components, and produce the residual images. The two-components are considered to be a central point source and extended disk component. These components are obtained by the 2D Gaussian fitting and subtracted on the image plane. Then, the residual images are expected to remain the disk substructures.
Figure \ref{residual} shows the images of the observations, the two-components models, and the residuals, respectively.

The substructure can be recognized in the residual image of the Band 6 data.
The peak emission is found with more than $5\sigma$ in the $\sim0\farcs4$ west from the protostar. 
In addition, the emission in the $\sim0\farcs5$ east is also found with $3\sigma$, suggesting that the local enhancements from the extended disk structure are remained in the both sides. 
Therefore, we suggest that the substructures are reliable features. 
If the emission of the both sides are real and symmetry, the ring structure is possible.
However, it is highly needed to investigate whether these substructures are ring-like or other lopsided structures in future studies with higher spatial resolutions.

Note that we cannot identify any substructures in the residual image of the Band 7 data. This is because the sensitivity of the Band 7 data are not enough to identify the substructure emission.

Although the substructure morphology is not conclusive in the Band 7 data, we check the spectral index in the $r\lesssim100$ au region.
Figure \ref{fig1} (f) shows the zoomed-in view of the spatial distribution of the spectral index $\alpha$. The distribution seems to show radial variations along the substructure.
A decrease of $\alpha$ may be found at the ring-like position.
Further observations with better spatial resolutions can assess the relation between the spectral index and substructure of the disk.

\begin{figure}[htbp]
\begin{center}
\includegraphics[width=8.cm,bb=0 0 2061 2556]{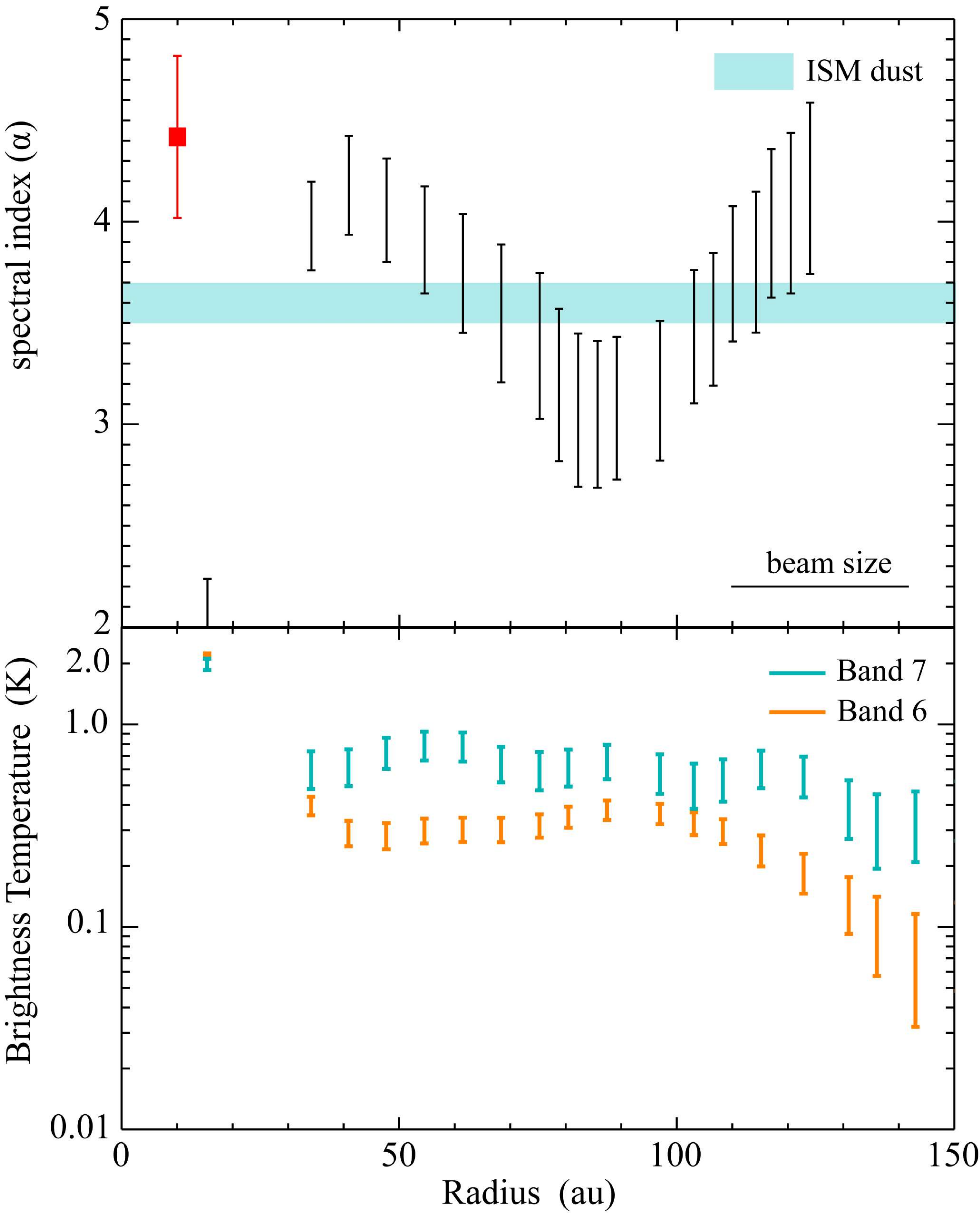}
\end{center}
\caption{Same radial profiles as Figure \ref{fig2} but zoomed-in views around the protostar shown in Figure \ref{fig1} (d-f).
The radial profiles are made toward the northeast direction with a position angles of $60^\circ$.
 The 1$\sigma$ error bars of the spectral index in Figure \ref{fig2} are derived from the image noise levels. In addition, the impact of a flux calibration
error of 10\% is shown by the red error bar.
}
\label{fig3}
\end{figure}

Figure \ref{fig3} shows the radial profiles of the spectral index and the brightness temperatures of Band 7 and 6 as similar to Figure \ref{fig2}.
The radial profiles are plotted only toward the northeast direction with a position angles of $60^\circ$ because the opposite side of the southwest direction has weak emission.
The substructure seems to be seen at a radius of  $r\sim80-90$ au and a width (full width at half maximum) of 80 au, derived by the gaussian fitting of the radial plot in the Band 6 emission. 
Note that the width of the ring structure may be overestimated because the gaussian fitting is performed with including not only the ring but also the extended disk component. 
The brightness temperatures show as low as 1 Kelvin in the 0.9 mm and 1.3 mm dust continuum emission even in the inner 100 au region, which suggests that the emission is optically thin around the substructure.
However, if the ring-like substructure is not resolved, the measured brightness temperature is underestimated and we cannot completely rule out that emission is optically thick.

Figure \ref{fig3} might imply that the spectral index seems to decrease around $r\sim 90$ au with $\alpha\sim3.0$.
The lowest spectral index of $\alpha\sim3.0$ is also found in the local peak position in the substructure toward the southwest direction shown in Figure \ref{fig1} (f).
These variations of the spectral index need to be studied by further observations with better spatial resolutions. 
The 1$\sigma$ error bars are derived from the image noise levels. In addition, the impact of a flux calibration
error of 10\% is shown by the red error bar.

\section{Discussion} \label{sec:dis}

\subsection{Spectral Index and Grain Size}

The spectral index and the intensity are related to the dust temperature ($T_{\rm d}$), the optical depth ($\tau_{\nu}$), and the dust opacity index ($\beta$).
The spectral index can be described as
\begin{equation}
\alpha=\frac{d\ln I_{\nu}}{d \ln \nu}=3-\frac{h\nu}{k_{\rm B}T_{\rm d}}\frac{e^{h\nu/k_{\rm B}T_{\rm d}}}{e^{h\nu/k_{\rm B}T_{\rm d}}-1}+\beta\frac{\tau_\nu}{e^{\tau_\nu}-1},
\label{eq1}
\end{equation}
where $h$ is Planck's constant, and $k_{\rm B}$ is Boltzmann's constant \citep{tsu16}.
If we use the Rayleigh-Jeans approximation and assume that the optical depth is much lower than unity, the equation (\ref{eq1}) becomes
\begin{equation}
\alpha\sim2+\beta.
\label{eq2}
\end{equation}
This approximation would work for the L1489 disk because the emission is optically thin.
Then, the spectral index $\alpha$ is linearly related to the dust opacity index $\beta$.
However, the Rayleigh-Jeans approximation may not be a suitable assumption in the outer disk region where the disk temperature decreases.
The disk temperature at $r\sim200$ au is suggested to be 30 K from the analysis of the spectral energy distribution \citep{bri07}.
In this temperature, the equation (\ref{eq2}) becomes $\alpha\sim1.76+\beta$.
In contrast, \cite{yen14} detected the SO emission in the disk with a radius of $250-390$ au.
Because the SO desorption temperature is $\sim60$ K, the temperature may be as high as 60 K. 
If we assume a disk temperature of 60 K, the equation (\ref{eq2}) becomes $\alpha\sim1.89+\beta$.
The different temperature causes an uncertainty of $\Delta\alpha\sim0.2$ due to the Rayleigh-Jeans approximation at $r\gtrsim200$.

\begin{figure}[htbp]
\begin{center}
\includegraphics[width=8.cm,bb=0 0 1031 698]{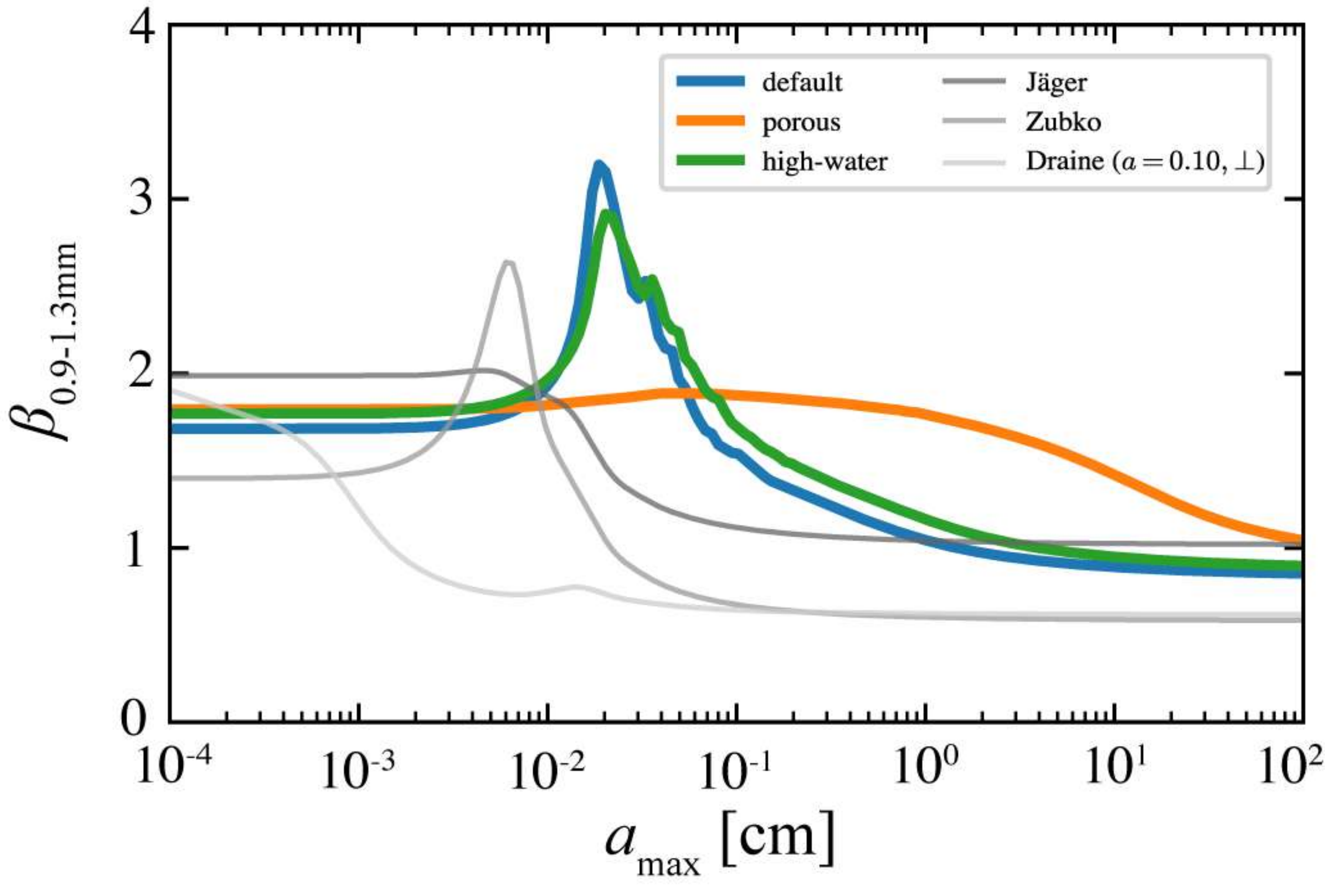}
\end{center}
\caption{The spectral index in the $0.9-1.3$ mm wavelength range with various dust models \citep{bir18}.
}
\label{fig4}
\end{figure}

The dust opacity index, $\beta$, is indicative of the grain size \citep[e.g.,][]{dra06,ric10,tes14}.
To investigate the grain size from the spectral index, we use the opacity model developed by \citet{bir18}.
Figure \ref{fig4} shows the dust opacity index $\beta$ modeled at $0.9-1.3$ mm wavelengths.
Even though $\beta$ has some variations depending on the dust models, Figure \ref{fig4} show a trend that $\beta$ decreases as increasing the grain size. 
The dust grains smaller than $a_{\rm max}\lesssim100$ $\mu$m has $\beta\sim1.5-2.0$.
The opacity model in DSHARP \citep{bir18} shows $\beta=1.68$ when the dust is small, which is shown as the blue line (default) in Figure \ref{fig4}.

 By taking the temperature uncertainty, the spectral index can be expected as $\alpha\sim3.5-3.7$ if the dust size is smaller than  $a_{\rm max}\lesssim100$ $\mu$m.
In  Figures \ref{fig2} and \ref{fig3}, the horizontal shadowed regions indicate $\alpha\sim3.5-3.7$. 
Due to the uncertainty of the spectral index, $\Delta\alpha=0.4$, the grain size cannot be constrained precisely.
However, according to those figures, the dust size is suggested to be the ISM-like dust (as small as $a_{\rm max}\lesssim100$ $\mu$m at $r\gtrsim100$ au), indicating that the grain growth does not occur significantly.
Even if we assume the lowest case within the error, the grain size needs to be smaller than $a_{\rm max}\lesssim1$ mm.
Therefore, we find no evidence of the grain growth in the outer part of the disk ($r\gtrsim100$ au), suggesting that the planet formation have not begun yet.

In contrast, Figure \ref{fig3} might imply that the dust size becomes as large as $a_{\rm max}\gtrsim1$ mm around the substructure of $r\sim90$ au because the spectral index seems to be lower $\alpha\sim3.0$.
Since the current observations are not enough to derive the spectral index, further observations with better sensitivity and spatial resolution are needed to measure the spectral index in the ring-like region precisely.

Note that the spectral index decreases inner the 30 au region. This is probably due to the high optical depth of the continuum emission, which follows the blackbody radiation with $\alpha=2.0$. However, it might also possible that larger dust grains are formed in the inner region because the brightness temperature is as low as $\sim2$ K.
Further observations with higher spatial resolution will also allow us to investigate the grain size in the central region.

\subsection{ No evidence of the grain growth but disk substructure formation}

Although the disk morphology is not clear, the tentative substructure is found around $r\sim90$ au.
If the substructure is real, one possibility of the substructure is a ring because the several protostellar disks with the ring-like substructures have been identified \cite[e.g.,][]{she20,seg20,alv20}.
Various mechanisms for the ring formation have been studied such as a growth front \citep{oha21}, snowlines of volatile gas species freezing-out onto dust grains \citep{zha15,oku16}, magneto-rotational instabilities \citep{flo15}, disk winds in unevenly ionized disks \citep{tak18}, unseen planets \citep{gol80,nel00,zhu12}, secular gravitational instability \citep{tak14}, and coagulation instability \citep{tom21}.

Our results show that the grain growth has not occurred yet outside of the ring-like position. This seems not to be consistent with some ring formation mechanisms that requires large dust grains such as the snow line model. In contrast, this result is consistent with the idea of the growth font, where the dust grains are grown by creating the ring structure. The grain growth proceeds in an inside-out manner, and the ring position corresponds to the growth front. Outside of the growth front, the dust particles remain in their initial state because they have not evolved yet, while the dust grains grow inside of the growth front.  The location of the growth front ($R_{\rm p}$) is estimated as 
\begin{equation}
R_{\rm p}=56\left(\frac{M_\star}{M_\odot}\right)^{1/3}\left(\frac{\zeta_{\rm d}}{0.01}\right)^{2/3}\left(\frac{t_{\rm disk}}{0.1\ {\rm Myr}}\right)^{2/3}\ {\rm au},
\label{eq:front}
\end{equation}
where $M_\star$, $\zeta_{\rm d}$,  $t_{\rm disk}$ are the protostellar mass, the gas to dust mass ration, and the disk age, respectively \citep{oha21}. 
By assuming that $M_\star=1.6$ $M_\odot$ \citep{bri07,yen13,yen14},  $\zeta_{\rm d}=0.01$, and $t_{\rm disk}=0.1$ Myr, the growth front is derived to be $\sim70$ au in the L1489 disk, which is consistent with the ring-like position.

If the ring-like structure is caused by the growth front, the disk inside the ring has grains as large as millimeter or even larger, implying the beginning of the planet formation.
The disk mass is estimated to be $\sim0.0071$ $M_{\odot}$ \citep{sai20}, which is a typical mass of the Class I protostellar disks \citep{tob20,she17}.  Therefore, the disk is still massive enough to form planets via accretion from the disk in this early stage \citep{kob20}.

Another mechanism of the ring formation is the presence of planets. 
In this case, planets need to be already existed in the Class I stage. 
Some other mechanisms of the ring formation such as magneto-rotational instabilities \citep[e.g.,][]{flo15} and disk winds \citep[e.g.,][]{tak18} may also work for the ring formation. 
It is highly needed to make clear the formation mechanism in future study.

\section{Summary} \label{sec:sum}

In this paper, we investigated the 0.9 and 1.3 mm dust continuum emission toward the L1489 protostellar disk. The dust continuum emission extends to be a disk radius of 300 au, and the spectral index is found to be $\alpha\sim3.6$ at a radius of $100-300$ au, as similar to the ISMs.
Therefore, the grain growth have not begun yet.
Furthermore, we tentatively dicovered the ring-like substructure around $r\sim90$ au in the 1.3 mm continuum image. 
If this is the real ring structure, the ring position and ISM-like dust grains are consistent with the growth front model, implying that the grain growth may occur inside the ring position.
These results suggest that the L1489 protostellar disk may be the beginning of the planet formation.
Further observations with better spatial resolutions allow us to reveal the disk morphology and constrain the formation mechanism of the disk substructure.

\acknowledgments

We gratefully appreciate the comments from the anonymous referee that significantly improved this article.
This paper makes use of the following ALMA data: ADS/JAO.ALMA\#2013.1.01086.S, and\\ ALMA\#2015.1.01549.S. ALMA is a partnership of ESO (representing its member states), NSF (USA) and NINS (Japan), together with NRC (Canada), MOST and ASIAA (Taiwan), and KASI (Republic of Korea), in cooperation with the Republic of Chile. The Joint ALMA Observatory is operated by ESO, AUI/NRAO and NAOJ. In addition, publications from NA authors must include the standard NRAO acknowledgement: The National Radio Astronomy Observatory is a facility of the National Science Foundation operated under cooperative agreement by Associated Universities, Inc.

This project is supported by pioneering project in RIKEN (Evolution of Matter in the Universe) and a Grant-in-Aid from Japan Society for the Promotion of Science (KAKENHI: Nos. 22H00179, 22H01278, 21K03642, 20H04612, 20H00182, 20K14533, 18H05436, 18H05438, 17H01103, 17H01105).

Data analysis was in part carried out on common use data analysis computer system at the Astronomy Data Center, ADC, of the National Astronomical Observatory of Japan.

\facilities{ALMA}

\software{CASA \citep[][]{mcm07}}

\end{document}